# Specific low frequency electromagnetic fields induce epigenetic and functional changes in U937 cells

Giulia Pinton[1], Angelo Ferraro[2], Massimo Balma[3] and Laura Moro[1]

[1] Department of Pharmaceutical Sciences, University of Piemonte Orientale, Novara, Italy;
[2] School of Electrical and Computer Engineering, National Technical University of Athens, Athens, Greece;
[3] ReD Lab, Ethidea Srl, Mathi (Turin), Italy;

Corresponding author E-mail: laura.moro@uniupo.it

**Abstract**

In this study, we investigated the effects of specific low frequency electromagnetic fields sequences on U937 cells, an in vitro model of human monocyte/macrophage differentiation. U937 cells were exposed to electromagnetic stimulation by means of the SynthéXer® system using two similar sequences, XR-BC31 and XR-BC31/F. Each sequence was a time series of twenty-nine wave segments. Here, we report that exposure (4 days, once a day) of U937 cells to the XR-BC31 setting, but not to the XR-BC31/F, resulted in increased expression of the histone demethylase KDM6B along with a global reduction in histone H3 lysine 27 (H3K27) tri-methylation. Furthermore, exposure to the XR-BC31 sequence induced differentiation of U937 cells towards a macrophage-like phenotype displaying a KDM6B dependent increase in expression and secretion of the anti-inflammatory interleukins (ILs), IL-10 and IL-4. Importantly, all the observed changes were highly dependent on the sequence's nature. Our results open a new way of interpretation for the effects of low frequency electromagnetic fields observed in vivo. Indeed, it is conceivable that a specific low frequency electromagnetic fields treatment may cause changes in chromatin accessibility and consequently in the expression of anti-inflammatory mediators and in cell differentiation.

## 1. Introduction

Low frequency Pulsed Electro Magnetic Fields (PEMF) have been found to produce a variety of beneficial effects and therefore successfully employed as adjunctive therapy for a wide range of clinical conditions [1-4]. Although properly configured electromagnetic signals demonstrate to regulate major cellular functions, including proliferation, differentiation and apoptosis [5-7], little is known about their biological mechanism of action.

In general, an electromagnetic radiation is generated when charged particles, as electrons, move through conductive materials. Such a radiation may diffuse in the surrounding environment and be absorbed by organic matter, including human cells, tissues and organs [8]. The electrical currents may occur as a constant flow of electrons, either in continuous or pulsed waves, resulting in the generation of an electromagnetic field whose intensity and characteristics are proportional to the applied electrical power and waveform.

Despite intense investigations carried out worldwide until now, it is not fully clear how low frequency electromagnetic fields affect the cellular physiology and the understanding about the role of wave parameters is still far away.

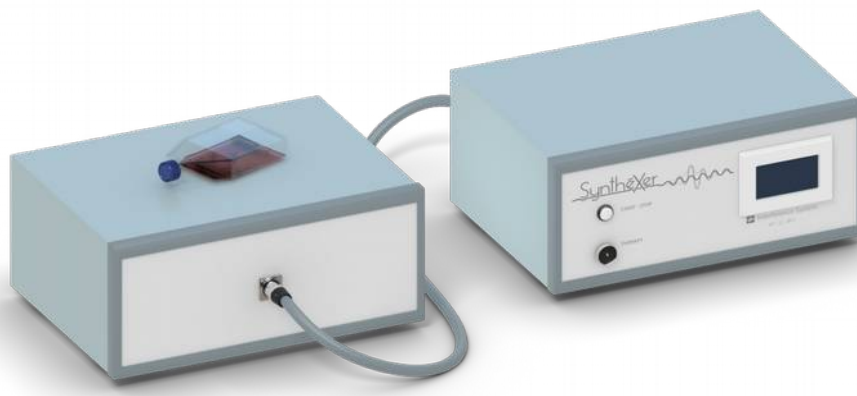

**Fig. 1** - A view of set-up for electromagnetic exposure tests. On the right side the SynthéXer® programmable generator, connected by cable to the antenna unit, on the left side. The test region is on the upper face of the antenna unit enclosure, where the cell culture flask is positioned.

From the biological point of view, it is well established that signals from the environment, whether they are physical, chemical, or hormonal, regulate intracellular metabolic processes such as enzyme activities and gene expression involved in cell differentiation and proliferation. Such metabolic mechanisms by which cells sense and respond to these signals are known as signal transduction mechanisms. Membrane signal transduction processes in particular have been an area of interest in research designed to elucidate effects of EMFs on cells.

Regardless of the large number of possible variables which can be sensed by signal transduction systems, there is a relatively limited number of mechanisms by which the information contained in these signals can be transmitted (transduced) across the cell membrane. In all known signal transduction systems, a molecule interacts with a protein located on the membrane (the receptor) and triggers conformational changes in the receptor which result in further modifications of cellular metabolism. Signalling agents which have limited ability to penetrate the cell membrane (e.g. peptide hormones, neurotransmitters and growth factors) interact with receptors which span the cell membrane. Interaction of the intracellular portion of the receptor with other intracellular (effector) molecules causes changes in the activities of cellular pathways [9].

For a living cell or tissue to respond functionally to an exogenous electromagnetic field, it is necessary that the radiation reaches and can be detected at appropriate molecular, cellular, or tissue site. Although still not completely elucidated, the mechanism of action of EMF signals at molecular and cellular level has been analyzed and a huge amount of literature strongly suggests ion or ligand binding in a regulatory cascade could be the signal transduction pathway [10–13].

One of the first models of interaction was created using a biophysical approach [10, 14, 15] in which an electrochemical model of the cell membrane was employed to theoretically predict a range of waveform parameters for which bioeffects might be expected. This approach was based on the assumption that voltage-dependent processes, such as ion or ligand binding and ion transport at, and across the electrified interface of the cell membrane, were the most likely EMF targets.

Several studies further developed this framework using Lorentz force considerations [16–20], and more specificity was hypothesized by including ion resonance and Zeeman–Stark effect [21]. These models suggested that combined actions from low-frequency alternating current (AC) and static magnetic fields as the geomagnetic one, could stimulate ion or ligand Larmor precession in a molecular binding site and thereby affect binding kinetics [22-27]. Direct action of the Lorentz force on free electrons in macromolecules such as DNA has also been proposed [28, 29].

At the present, the most accepted, basic biophysical transduction step is ion or ligand binding at cell surfaces and junctions, which modulate a cascade of biochemical processes resulting in the observed physiological effect [24-27, 30]. Several theoretical models have been formulated to quantify, by practical calculations, the effects of field interaction. Among them, it has been proposed a detailed physical interpretation of the forced-vibration of free ions present on the external surface of the plasma membrane, which in the end changes the cell electrochemical balance and function [31,32].

Not long ago, new insights on the biological role of electromagnetic interactions also arise from the molecular investigation on endogenous bioelectric signals during pattern formation in growing tissues [33-35]. As mentioned before, ion flows and voltage gradients produced by ion channels and pumps are key regulators of cell proliferation, migration, and differentiation. In addition, thanks to the use of fluorescent voltage reporters and functional experiments using well-characterized channel mutants, instructive roles for bioelectrical gradients in embryogenesis, regeneration, and neoplasm are being revealed [34]. This growing knowledge trend, therefore, will fully integrate bioelectric signalling pathways with known genetic and biochemical cascades.

Beside insight concerning the general models, although elementary, on how electromagnetic field may affect cell membrane tasks, accumulating data suggest that PEMF treatments lead to changes in the expression of genes involved in regulating inflammation, including inflammation resolution [36]. Indeed, exposure of human peripheral blood mononuclear cells to PEMF results in modified expression of a number of cytokines and in particular interleukins, which are able to modulate the immune and inflammatory responses [37-41].

Interleukins (ILs) are a broad category of small proteins generally grouped in two families based on their ability to promote (pro-inflammatory ILs) or to block the inflammation cascade (anti-inflammatory ILs). Even though a clear division in pro- and anti-inflammatory ILs it is not always possible, because of dual roles played by many of those molecules, it is accepted that some cytokines such as IL-1 and tumor necrosis factor (TNF) show pro-inflammatory activity [42], whereas others such as IL-4, IL-6, IL-10, IL-11 show anti-inflammatory activity [43]. In particular, IL-4 alone or in association with IL-13, has been described to cause polarization of macrophages into an anti-inflammatory or M2 phenotype. M2 macrophages exhibit reduced pro-inflammatory cytokine secretion, produce anti-inflammatory factors (e.g. IL-10) which dampen inflammatory and adaptive Th1 responses, and show high levels of scavenger, mannose, and galactose-type receptors, which contribute to tissue repair [44]. The process of macrophage polarization involves an intricate interplay between various cytokines, chemokines, transcriptional factors, and immune-regulatory cells [45]. Besides humoral dynamics, increased attention has been given to epigenetic changes that affect macrophage functional responses and M1/M2 polarization, through modulating gene expression signature.

KDM6B, also known as JMJD3, is a chromatin remodeling factor which specifically demethylates di- or tri-methylated lysine 27 of histone H3 (H3K27me2 or H3K27me3), affecting gene-expression [46, 47]. Post-translational modifications of histone tails, and in particular methylation, deeply affect the overall chromatin structure during crucial cellular processes. Indeed, H3K27me3 is enriched in highly condensed and thereby inactive chromatin, whereas loss of H3K27me3 is linked with transcriptionally active genes [48-50]. The action of KDM6B is counteracted by EZH2, the catalytic subunit of the polycomb repressor complex 2 (PRC2), whose primary function is to add methyl groups at lysine 27 on histone H3 by using the cofactor S-adenosyl-L-methionine (SAM) [51].

In this paper, we provide evidence that cellular responses to low frequency electromagnetic fields are highly dependent on the nature of administrated electromagnetic waves.

Expression and activity of KDM6B, as well as the release of specific anti-inflammatory interleukins in U937 cells when exposed to complex electromagnetic fields sequences, have been analyzed. We report that certain time patterned wave groups, of suitable and narrow frequency windows, are capable to induce or not evident metabolic and phenotypic differences in the exposed cells with respect to the control ones.

Our experimental observations suggest new mechanisms which allow widening, and perhaps overcoming the biophysical approach based on coherent vibration of ions charges.

## 2. Results

Following the setup of Fig.1, cells were exposed 4 days, once a day for 77 minutes, to electromagnetic fields by means of SynthéXer® system (Ethidea Srl, Turin, Italy), an electronic device able to generate and deliver electro-magnetic signals with arbitrary waveform, in the frequency range from 10 Hz to 10 kHz.

### 2.1. Description of the stimulus sequence

The general frame of SynthéXer® signal is based on a train of N wave segments that has been referred to as specific time sequence S(t):

$$S(t) = \sum_{i=1}^{N} A_i X_i(t) \, Rect\left[\frac{t - \sum_{p=1}^{i-1} \Delta_p + \Delta_i/2}{\Delta_i}\right]$$

where $A_i$ is the amplitude of the signal into the segment i, $X_i(t)$ represents the i-th waveform with its relevant parameters and $\Delta i$ is the time duration of the i-th segment. Within each segment, several well defined parameters such as frequency, waveform, modulation and level were therefore applied, as shown in Figures 2 and Figure 3. In the experiments herein reported, a SynthéXer® sequence, named XR-BC31, was used. The XR-BC31 sequence was formed by a series of twenty-nine wave segments, equal to a total duration of 77 minutes. The waveform in every segment was a monochromatic square wave with 50% of duty cycle and specific frequency. The range of generated frequencies into the sequence varied from about 20 Hz to 3200 Hz (Figure 2).

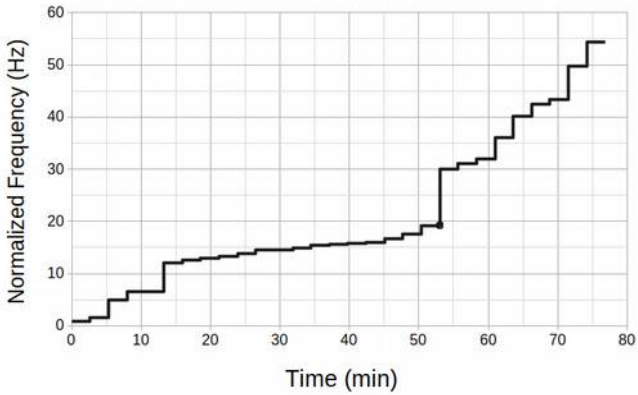

**Fig. 2** - Frequencies of square waves $X_i(t)$ as function of time along the XR-BC31 sequence.

Before beginning the experiment and periodically during the tests, the magnetic flux density, as emitted from the device into the test region, was measured with a calibrated Electromagnetic Field Analyzer for monitoring low frequency fields (EFA-300, Narda Safety Test Solutions, www.nardasts.com).

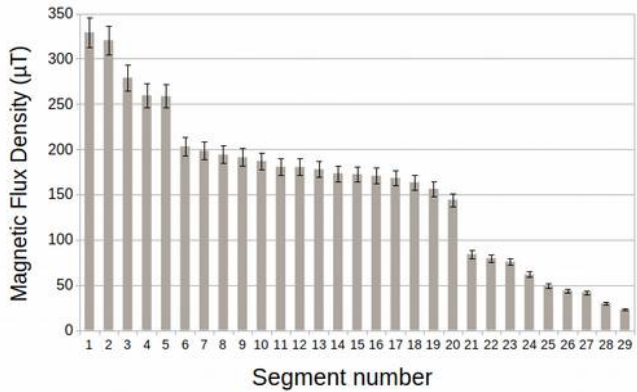

**Fig. 3** - Magnetic flux measured on the test region as function of XR-BC31 segments.

Stray ambient AC fields were below 0.18 μT. The measured XR-BC31 magnetic flux emissions into the test region ranged from about 20 uT to 400 μT, over the whole operative wave segments (Figure 3).

Instantaneous $W_M$ energy associated with those magnetic field levels can be expressed as a function of magnetic flux density (B), volume (Vol) and magnetic permeability, $\mu_0 \mu_R$ as follows:

$$W_M = \frac{1}{2\mu_0\mu_R}|B^2|(Vol)$$

where $\mu_0 = 4\pi \times 10^{-7}$ H/m and $\mu_R \approx 1$. Therefore, considering the volume of a cell of 10 μm radius, the range of energies associated with 20 μT to 400 μT magnetic flux of XR-BC31 is approximately equivalent to $1.3 \times 10^{-18}$ J at $0.51 \times 10^{-15}$ J which is from 2 to 5 orders of magnitude larger than thermal fluctuation. As previously reported, an electric field *E* may also be set up in tissues and cells by a time-varying magnetic field according to Faraday's law. Such electric field *E* is induced as:

$$E = \omega r_C B/2$$

where ω is the field angular frequency, $r_C$ the cell radius and B the value of magnetic field. The peak values of this magnetically induced electrical field *E*, at the frequencies and magnetic flux generated with XR-BC31 sequence, fell in the μV/m range.

## 2.2. Exposure to the XR-BC31 sequence induced changes in U937 morphology and cell cycle distribution

We analyzed the effects of exposure to low frequency electromagnetic fields, generated by means of SynthéXer® system, on the human promonocytic U937 cell line.

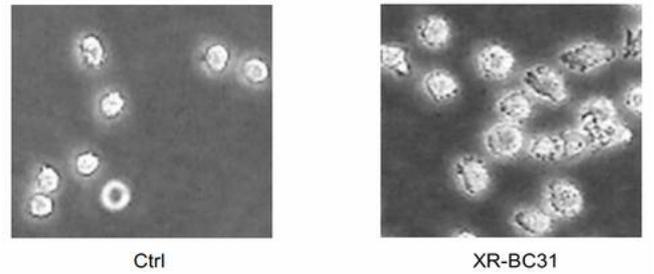

**Fig. 4** - Phase contrast microphotographs (400 X magnification) of U937 cells exposed 4 days (once a day) (XR-BC31) or not (Ctrl) to the XR-BC31 sequence.

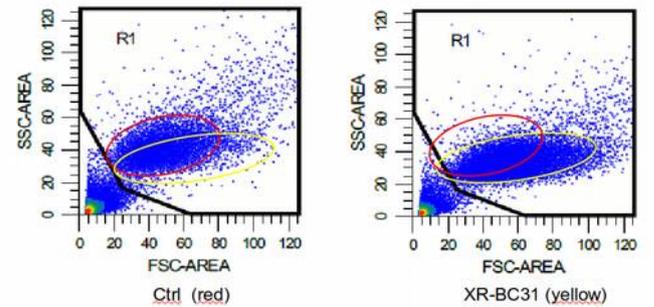

**Fig. 5** - Flow cytometry dot-plots showing forward versus side-scatter profiles of control cells (circled in red) and cells exposed to XR-BC31 sequence (circled in yellow).

Exposure of U937 cells to the XR-BC31 sequence (4 days, once a day) resulted in morphological and functional changes. As evidenced by microscope analysis (Figure 4), U937 cells which normally grow in suspension and show a smooth surface, after treatment extended pseudopodia and became more adherent acquiring a macrophage- or dendritic-like morphology.

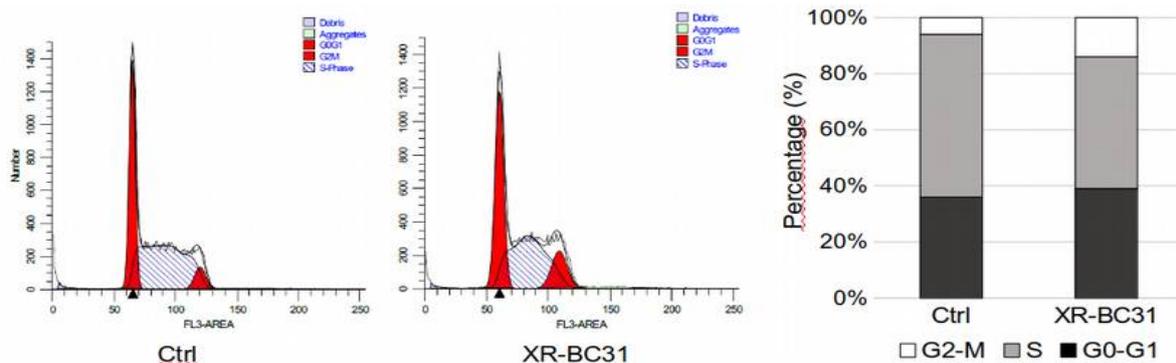

**Fig. 6-** Cell cycle analysis of control cells (Ctrl) and cells exposed (XR-BC31) during 4 days (once a day) to the SynthéXer XR-BC31 sequence. After treatments, cells were stained with propidium iodide and analyzed for cellular DNA content by flow cytometry. Exemplificative histograms that plot cell count versus DNA content and bar graphs showing the quantification of each cell phase are reported.

Cytofluorimetric analysis evidenced changes in size and complexity/granularity in exposed cells (gated in yellow) compared to control cells (gated in red) (Figure 5). Furthermore, concomitant with morphological changes, flow cytometric analysis of cell cycle distribution evidenced that U937 cells exposed to the XR-BC31 sequence accumulated in G2/M phase (Figure 6).

### 2.3. Exposure to the XR-BC31 sequence induced expression and secretion of IL-4 and IL10 in U937 cells

To investigate the effects of electromagnetic fields on cytokines release, we detected cytokines in culture supernatants of U937 cells exposed 4 days (once a day) to the XRBC31 sequence, using a multi-cytokine ELISA kit.

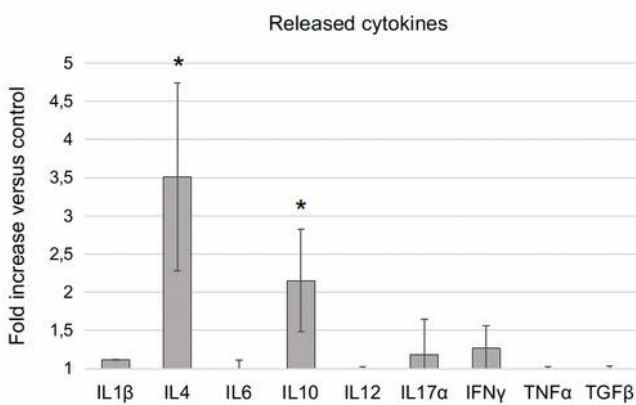

**Fig. 7** - Bar graph showing cytokines release in culture supernatants of U937 cells exposed 4 days (once a day) to the XR-BC31 sequence, as determined by multi-cytokine ELISA assay. Each bar represents mean of three independent experiments ± SD. * p≤0.05.

The grap in Figure 7 shows that the release of the anti-inflammatory cytokines, IL-4 and IL-10, increased significantly in U937 cells exposed to the XR-BC31 sequence, when compared to control cells.

By contrast, we did not detect significant differences in IL-1β, IL-6, IL-12, IL-17α, IFNγ, TNFα and TGFβ 6 release. Furthermore, we did not observe significant induction in cytokine release in supernatants of cell cultures exposed to a single stimulation run with the XR-BC31 sequence, either recovered immediately after exposure or after four days of culture (not shown). In accordance with the observed protein release, we demonstrated that 4 days (once a day) exposure of U937 cells to the XR-BC31 sequence significantly induced IL4 and IL10 gene transcription (Figure 8).

### 2.4. Exposure to the XR-BC31 sequence induced KDM6B expression and H3K27 demethylation in U937 cells

Well-established literature has demonstrated that low frequency/low energy electromagnetic fields do not cause predictable effects on DNA in terms of mutations or any other chemical change. Nevertheless, actuation of a genetic program is not solely controlled by DNA sequence changes, but rather also by epigenetic mechanisms on which electromagnetic fields may act.

We performed experiments to study the effect of low frequency electromagnetic fields exposure on expression of the two epigenetic modifiers KDM6B and EZH2.

As shown in Figure 9 and Figure 10A, exposure to the XR-BC31 sequence significantly induced KDM6B expression in U937 cells, while the level of EZH2 remained unchanged.

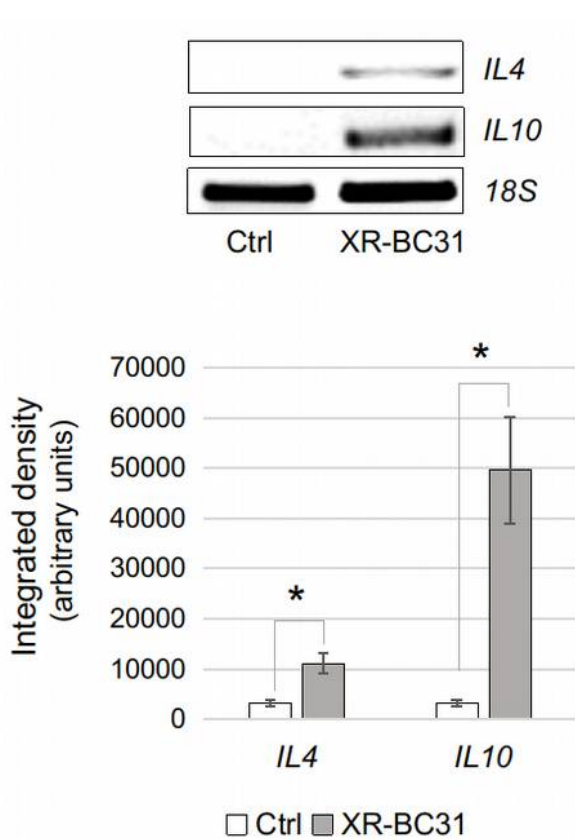

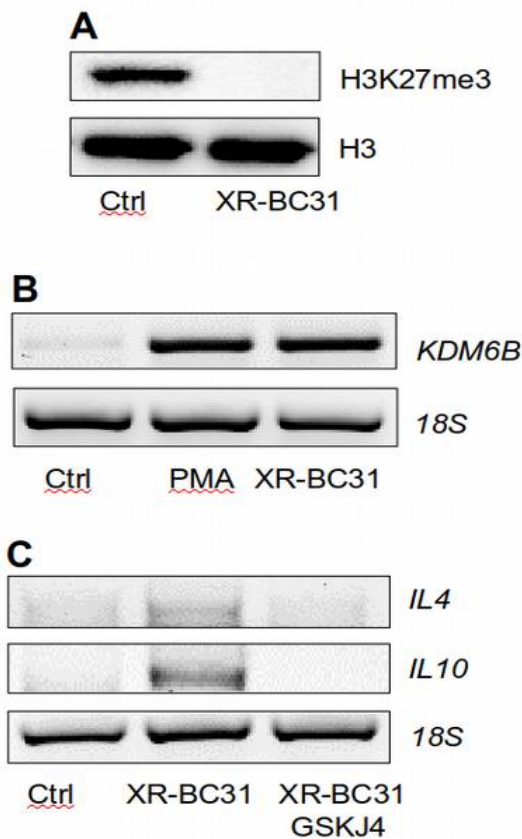

**Fig. 8** - Representative RT-PCR and densitometric analysis of IL4 and IL10 expression (cropped images are parts of the same gel) in control (Ctrl) and XR-BC31 sequence exposed U937 cells (XR-BC31). 18S was used as housekeeping gene. In the graph, each bar represents mean of three independent experiments. * p≤0.05

**Fig. 10** - (**A**) Representative Western blot analysis of H3K27me3 and histone H3 in U937 cells exposed 4 days to the XR-BC31 sequence (once a day), by means of the SynthéXer® system. (**B**) Representative RT-PCR analysis of KDM6B expression in U937 cells untreated (Ctrl), treated 48 hours with 10 nM PMA (PMA) or exposed 4 days to the XR-BC31 sequences. (**C**) Representative RTPCR analysis of IL4 and IL10 expression in U937 cells not exposed (Ctrl) or exposed to XRBC31 sequence in the absence or in the presence of the selective KDM6B inhibitor, GSKJ4. 18S was used as housekeeping gene.

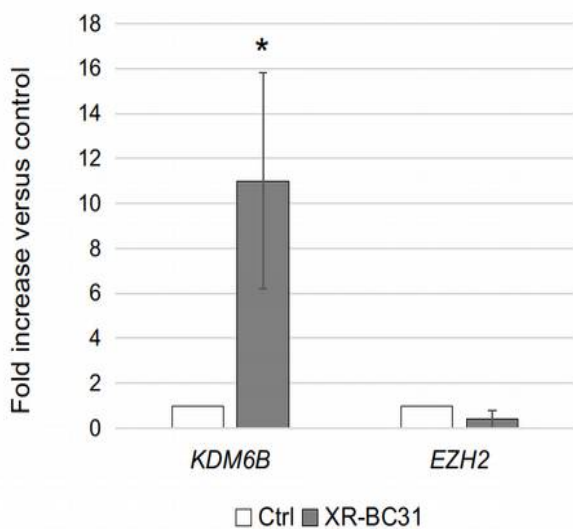

**Fig. 9** - PCR analysis of KDM6B and EZH2 expression in U937 cells exposed 4 days to XR-BC31 sequence (once a day), by means of the SynthéXer® system. Each bar represents mean of three independent experiments. * p≤0.05

Similar increase in KDM6B expression was observed when U937 cells were stimulated 48 hours with 10 nM PMA, a well-studied differentiation-inducing chemical agent (Figure 10B).

Notably, IL4 and IL10 transcription was not induced when cells were exposed to the XR-BC31 sequence and concurrently treated with the KDM6B selective inhibitor, GSKJ4 (Figure 10C).

### 2.5. Exposure to the XR-BC31/F sequence did not induce KDM6B and ILs expression in U937 cells

On the baseline of obtained results, another series of experiments were carried out with the aim to appreciate the specificity of sequence parameters. To this purpose, a second SynthéXer® sequence was designed and named XR-BC31/F. Starting from the general assumption that magnetic field interaction efficacy should be frequency related to cellular phenomena timing, the only parameter variation that was introduced into XR-BC31/F was the frequency of square waves $X_i(t)$. Figure 11 shows frequencies of segment square waves as function of time.

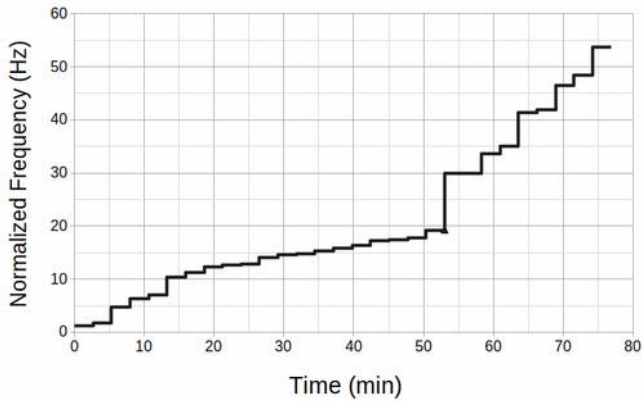

**Fig. 11** - Frequencies of square waves $X_i(t)$ as function of time along the XRBC31/F sequence, again composed by a series of 29 wave segments.

The values were compared with those reported in Figure 2 and the percentage difference (Freq$_i$ XR-BC31/F - Freq$_i$ XR-BC31) / Freq$_i$ XR-BC31, where the apex i identifies the *i-th* segment, is reported in Figure 12.

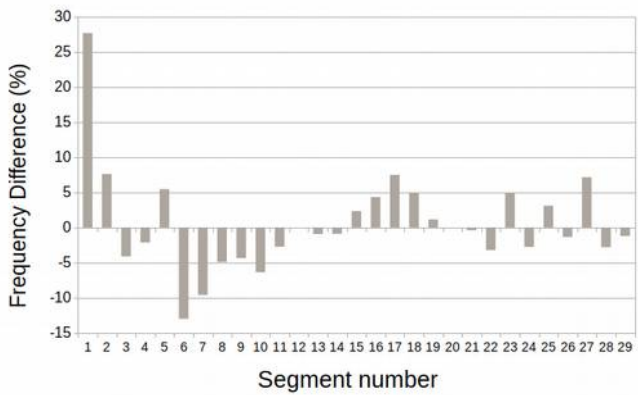

**Fig. 12** - Differences of square waves $X_i(t)$ frequencies as function of segment number, with respect to XR-BC31.

The absolute value frequency deviations (in percentage) was chosen on a random manner, with an average value of 4.72 ± 5.35.

As shown in Figure 12, the frequency deviations with respect to the XR-BC31 baseline were both positive and negative, and even nil for some wave segments.

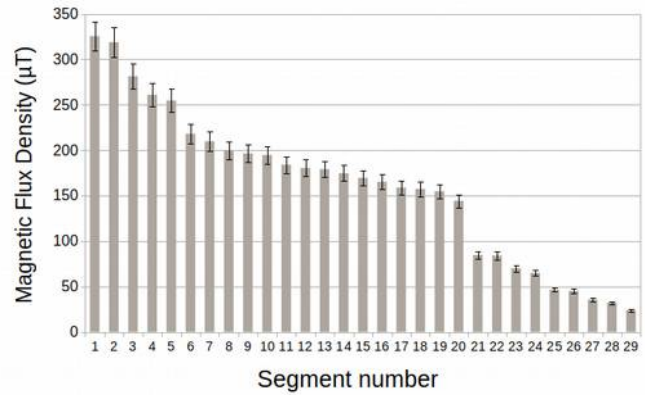

**Fig. 13** - Magnetic flux measured on the test region as function of XR-BC31/F segments.

As previously reported, also for the XR-BC31/F magnetic flux emissions into the test region was measured and ranged, over the whole operative wave segments, from about 20 uT to 400 uT (Figure 13). The global energy content into the two sequences was reasonably the same because frequency deviations were small and measured field levels were very close.

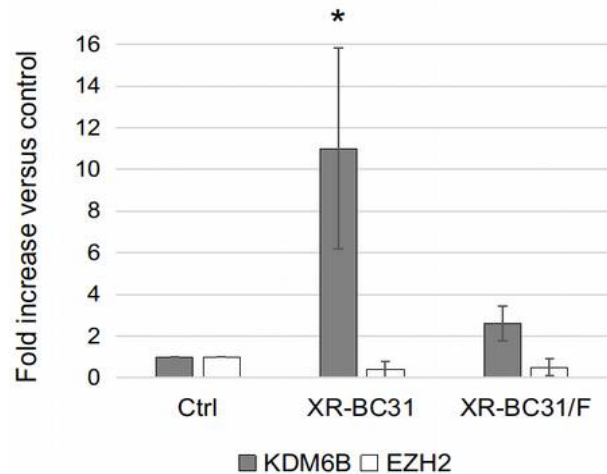

**Fig. 14** - Real time PCR analysis of KDM6B and EZH2 expression in U937 cells exposed 4 days to the XR-BC31 or the XR-BC31/F sequence (once a day), by means of the SynthéXer® system. Each bar represents mean of three independent experiments. Significance is for p≤0.05

Importantly, by performing exactly the same experiments, we did not observe significant changes in KDM6B expression (Figure 14) and in global level of H3K27me3 (Figure 15A) in U937 cells exposed to the XR-BC31/F sequence when compared with cells exposed to the XR-BC31. Furthermore, in cells exposed to the XR-BC31/F sequence, we did not observe increased IL4 and IL10 gene transcription (Figure 15B).

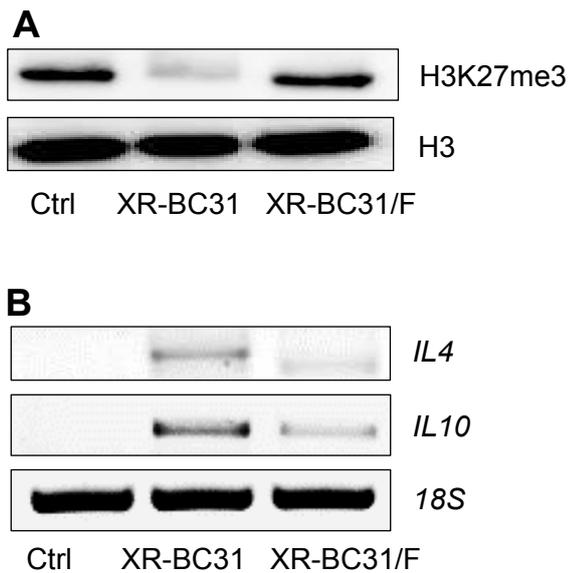

**Fig. 15** - (**A**) Representative Western blot analysis of H3K27me3 and histone H3 in U937 cells exposed 4 days to the XR-BC31 or the XR-BC31/F sequence (once a day), by means of the SynthéXer® system. (**B**) Representative RT-PCR analysis of IL4 and IL10 expression in U937 cells not exposed (C) or exposed to the XR-BC31 or the XR-BC31/F sequence. 18S was used as housekeeping gene.

## 3. Discussion

PEMF stimulation therapies have been clinically successful in treating a variety of medical conditions, including expedited healing of bone fractures, increased fusion rate for cervical and lumbar fusions, and arrested progression of osteoarthritis [52]. Although some clinical benefits of PEMF have been demonstrated, their efficacy is still debated and obfuscated by contradictory studies in which various PEMF settings have been used by investigators. A broad consensus exists on the fact that an applied electric field or an induced electric field associated with an applied magnetic field can influence a number of cell-membrane processes. Within this framework of signals, field-triggered changes occurring on the plasma membrane, such as ion flux and receptor binding events, are expected to translate into changes in mRNA and protein expression. Further studies are needed to better characterize the mechanisms by which PEMF modifies cell behavior and to identify cell markers related to such exposure. It has been described that PEMF treatment significantly affects the expression of genes associated with early stages of inflammation [53]. In addition, environmental signals at the site of inflammation mediate rapid monocyte mobilization and dictate differentiation programs whereby these cells give rise to macrophages or dendritic cells [54].

The present study was aimed to investigate the effects of specific low frequency electromagnetic sequences on monocytes differentiation. To address this issue, we used a commercial PEMF generator named SynthéXer®, which has been approved for the treatment of osteoarticular pathologies. The administration protocol of electromagnetic fields to the cell culture was similar, in terms of type of sequences and time of exposure, to those used to treat patients.

For the sake of comparison, while the minimum value of applied field to which a cell would respond without any selective, cooperative or amplifying mechanisms, was calculated to be about 1mV/cm for a large elongated cell and 20-40 mV/cm for a spherical cell [55], the peak values of the magnetically induced electrical field in our experimental setup fell in the μV/m range. On the other hand, it is known that some animals possess an extreme sensitivity to electric field and to magnetic fields [56].

As cell model, we used the human monocyte-like histiocytic lymphoma U937 cell line. Once stimulated by phorbol 12-myristate 13-acetate (PMA), U937 cells differentiate into macrophage-like cells, changing from non-adherent to adherent cells with extended pseudopodia. Hence, thanks to this biological setup, it is possible to study molecular changes linked to differentiation both as result of chemical stimulation and as result of physical stimulation by PEMF. Furthermore, because primary tissue macrophages cannot be readily expanded ex vivo, differentiation of monocytes from primary leukemia cell lines into macrophage-like cells has been frequently used as a mimic model for understanding the process of innate and adaptive immune responses to inflammatory stimuli [57, 58].

Analysis by phase-contrast microscopy and flow cytometry revealed changes in morphology, adhesive and growth properties, towards a macrophage-like phenotype, when monocytic U937 cells were exposed to the XR-BC31 sequence.

Epigenetic reprogramming is thought to play an important role during monocyte differentiation and macrophage polarization [59]. Previous studies have reported that KDM6B, a H3K27me2/3-specific demethylase, provides a supportive molecular activity that dictates in macrophages the M2 gene-expression profile. Indeed, KDM6B, specifically induced by IL-4, is recruited to the promoter regions of M2 marker genes [60].

To verify the involvement of epigenetic mechanisms in our model of monocyte differentiation induced by low frequency electromagnetic fields, we analysed and compared the expression of KDM6B and EZH2 in U937 cells in three conditions: cells exposed to the XR-BC31 sequence, cells exposed to the XR-BC31/F sequence and cells exposed to no one sequence (Ctrl). We found increased expression of KDM6B in cells exposed to the XR-BC31, but not to the XR-BC31/F sequence, while, EZH2 expression remained constant in both conditions. The observed global reduction in H3K27 tri-methylation confirmed increased KDM6B activity on chromatin along the whole U937 cells' genome.

The relevant finding for a crucial involvement of KDM6B in U937 cells differentiation was supported by the observed increase in KDM6B expression when cells were chemically induced to differentiate by PMA treatment.

These results suggest that the functional changes observed during U937 cell differentiation are due to epigenetic reprogramming that can be activated either via electromagnetic stimulus or via a chemical agent.

Macrophages express a battery of bioactive molecules that promote tissue remodeling/healing, support cell proliferation and angiogenesis, and mediate immunosuppression under certain micro-environmental conditions [61]. To further prove the activation of a differentiation program, we evaluated the release of several cytokines in culture supernatants of U937 cells stimulated or not with electromagnetic fields. The results showed that IL-4 and IL-10 release increased in the growth media of cells exposed four days to the XR-BC31 sequence. By RT-PCR, we evidenced a remarkable increase also in the expression of IL4 and IL10 mRNAs.

The increased expression and secretion of IL-4 in U937 cells exposed to XR-BC31 sequence was unexpected and needs further investigation. IL-4 release might be induced in cells through autocrine and paracrine mechanisms, which in turn could contribute to the acquisition of the differentiated phenotype.

IL-10 is a 35 kD cytokine produced by activated macrophages, B and T cells [62]. IL-10 is a pleiotropic cytokine with anti-inflammatory and immunoregulatory functions that plays a critical role in the containment and eventual termination of the inflammatory response [63]. IL-10 therapy is effective in a number of animal models of arthritis [64]. Notably, it has been reported that local expression of IL-10 at the inflammatory site has undoubtedly proven more effective than systemic administration in alleviating disease, due to its short half-life.

The possibility to induce macrophages to release IL-10 via EMF stimulation in vitro could be considered as groundwork to support in vivo anti-inflammatory mechanism of action for this therapy.

It is worth noting that the described biological effects were not recorded when the XRBC31/F sequence was used to stimulate cells or when stimulation was performed in the presence of a KDM6B selective inhibitor. Moreover, cell differentiation was not detectable after the first exposure, but after 4 days, one exposure at day, indicating that an epigenetic reprogramming and/or accumulation of secreted factors in cell growth media were required.

That a particular low frequency electromagnetic sequence could or not exert a biological effect, which can be framed in an anti-inflammatory action, is extremely intriguing. Indeed, during the past decades, several theoretical models have been carried out trying to introduce weighed or specific parameters into the electromagnetic signals with the aim of obtaining targeted effects [65]. Experimental evidences that specific frequencies modulate cellular functions have also been supported by the concept that ion cyclotron magnetic resonance has a role in regulating biological information [66, 67]. In this study we extended the possible ways of interpretation and report for the first time detailed parameters of electromagnetic signals that, even though very similar, are able to give significant different effects on a living biological system. The parameters that we used to design two different sequences were the segment frequency values of square waves Xi(t), with an average change of 4.72%.

The observational meaning of this behaviour is that electromagnetic fields transfer instructions to the cells, whose efficacy seems reasonably related to the timing and triggering of cellular dynamics phenomena, within a narrow frequency bandwidth lower than 5%.

From the observed increased expression and activity of KDM6B is conceivable that a specific low frequency electromagnetic fields treatment can cause changes in the chromatin accessibility and consequently in the expression of mediators of the inflammatory response and cell differentiation. Nonetheless, the understanding of clear rules about the electromagnetic manner of communicating in biology is still far away.

Future research must address the molecular targets of such interactions and their triggering time. Therefore, by testing several wave parameters it would be feasible to discover certain electromagnetic fields sequences that could stimulate or inhibit given signalling pathways and cellular activities. Once confirmed, a conscious use of these selective sequences might provide an important, economical, non-invasive and non-pharmacological adjunct therapy.

## 4. Methods

### 4.1. Reagents and antibodies

The polyclonal antibodies specific for histone H3 trimethyl K27 (H3K27me3) and histone H3 were purchased from Santa Cruz Biotechnology (Santa Cruz CA, USA).

Anti-rabbit IgG peroxidase conjugated antibodies, Phorbol 12-Myristate 13-Acetate (PMA) and other common reagents were from Sigma-Aldrich (St Louis, MO, USA).

ECL, nitrocellulose membranes and protein assay kit were from Bio-Rad (Hercules, CA, USA). TRIzol®, culture media and sera were from Thermo Fisher (Waltham, MA, USA).

The KDM6B selective inhibitor, GSKJ4, was supplied by Selleckchem (Houston, TX, USA).

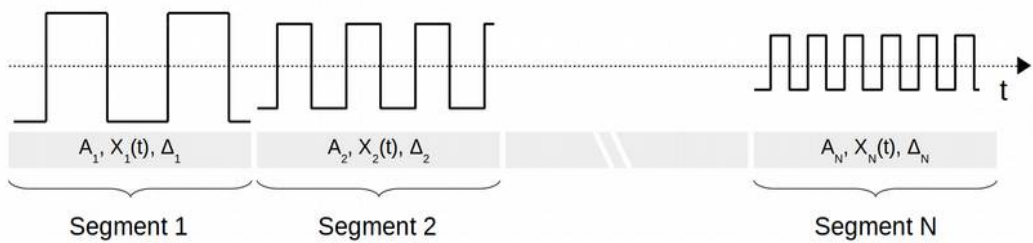

**Fig. 16** - The general frame of SynthéXer® signal is based on a series of N wave segments, where every segment holds a specific duration time. Within each segment several technical parameters such as frequency, waveform $X_i$, modulation and amplitude $A_i$ are defined.

### 4.2. Cell culture and low frequency electromagnetic stimulation

The U937 cell line was purchased from Sigma-Aldrich (St Louis, MO, USA). Cells were grown in standard conditions in RPMI medium supplemented with 10% FBS, 100 μg/ml streptomycin and 10 μg/ml penicillin at 37°± 0.5° in a humidified environment containing 5% controlled CO2.

Mycoplasma infection was excluded by the use of Mycoplasma PlusTM PCR Primer Set kit from Stratagene (La Jolla, CA, USA).

Cells were exposed 4 days, once a day for 77 minutes, to electromagnetic fields by means of the SynthéXer® system, adapted for in vitro experiments (Figure 1).

SynthéXer® E01S01025-01 (Ethidea Srl, Turin, Italy) is a medical device in Class IIa, able to generate and deliver programmable signals with arbitrary waveform in the frequency range from 10 Hz to 10 kHz (Figure 16). Its intended use is PEMF therapy for the treatment of bone fractures healing and osteoarticular problems in general, of osteoporosis and tendinopathy, inflammation of nerves and tissues, and pain affecting the musculoskeletal system.

In order to expose cells culture to electromagnetic sequence, a rectangular magnetic antenna was designed by means of FEM code (COMSOL Multiphysics®, www.comsol.it). The antenna was realized by a multiturn, single layer, copper wire solenoid 33 cm wide and 25 cm high, and was framed into the inner part of a plastic enclosure. The upper side of such antenna enclosure act as support for the flask containing the cells undergoing electromagnetic exposure. The solenoid into the enclosure is placed with its axis vertical so that the magnetic flux is perpendicular to the horizontal faces of the rectangular enclosure itself and it is driven by SynthéXer® output signal by means of a double shielded cable (Figure 17).

The "test region" where the flasks are positioned is the central part of the upper face of rectangular enclosure, where the field distribution is reasonably uniform.

The antenna is then located within an incubator with controlled temperature (37°± 0.5°) and 5% CO2. (Figure 18). The exposure tests were carried out on cells samples following different paths: one sample was placed on the test region, and it was exposed to the defined sequence (SynthéXer® ON). The corresponding sham exposed sample was placed under the same logistic and electrical conditions but with the solenoid radiating no field (SynthéXer® OFF).

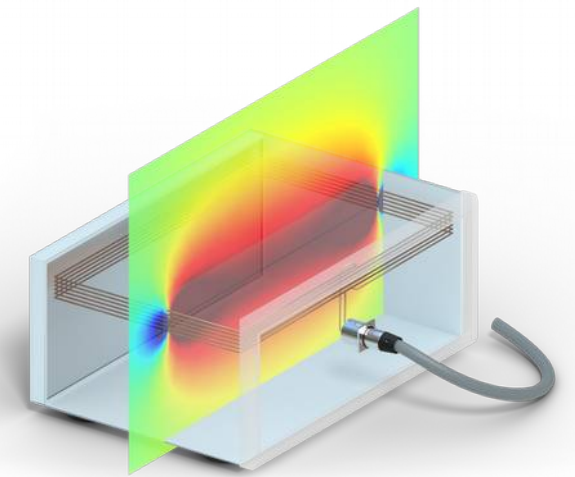

**Fig. 17** - The antenna, made by a rectangular and balanced solenoid, is housed in a plastic enclosure. The contour plot, simulated with COMSOL Multiphysics®, software, shows the computed levels of the magnetic field in the longitudinal-vertical plane.

Before beginning the experiment and periodically during the tests, the magnetic flux density, as emitted from the device into the test region, was measured with a calibrated Electromagnetic Field Analyzer for monitoring low frequency fields (EFA-300, Narda Safety Test Solutions, www.nardasts.com).

### *4.3. Cell cycle analysis*

For cell cycle/apoptosis analysis, $5 \times 10^5$ cells were exposed 4 days to treatment with SynthéXer® sequence XR-BC31, once a day. After treatment, cells were harvested in complete RPMI and centrifuged at 500 x g for 10 minutes. Pellets were washed with PBS, fixed in ice-cold 75% ethanol at 4° C, treated with 100 mg/mL RNAse A for 1 hour at 37° C, stained with 25 μg/mL propidium iodide and finally analyzed by using a S3e™ Cell Sorter 12 (Bio-Rad, Hercules, CA, USA) and Modfit software (Verity Software House, Topsham, ME, USA).

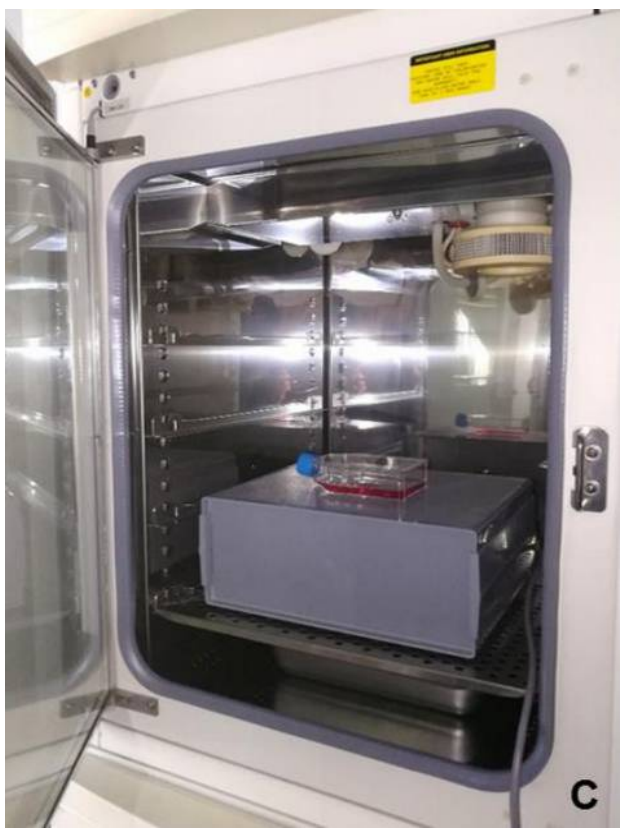

**Fig. 18** - Photo of the antenna enclosure housed inside a 37°C, CO2 controlled incubator for experiments.

### *4.4. ELISA for cytokines measure*

The simultaneous detection of released cytokines in medium samples was performed using the Multi-Analyte ELISArray Kit-336161 from Qiagen (Hilden, Germany) according to manufacturer instructions.

### *4.5. Cell lysis and immunoblot*

Histones were acid extracted from nuclei with 0.4 N HCl and precipitated with trichloroacetic acid (TCA), followed by washing with ice-cold acetone containing 0.006 % HCl, and then with pure ice-cold acetone. The resulting pellets were air-dried, dissolved in a minimal volume of sterile distilled water and the protein concentration was determined with the BioRad protein assay method. Proteins were separated by SDS-PAGE under reducing conditions. Following SDS-PAGE, proteins were transferred to nitrocellulose, reacted with specific antibodies and then detected with peroxidase-conjugate secondary antibodies and chemioluminescent ECL reagent.

Digital images were taken with the Bio-Rad ChemiDocTM Touch Imaging System and quantified using Bio-Rad Image Lab 5.2.1.

### *4.6. RNA isolation, semi-quantitative RT-PCR and Real Time PCR*

Total RNA was extracted using TRIzol® reagent. Starting from equal amounts of RNA (5 μg) cDNA was synthesized by the reverse transcription reaction using RevertAid Minus First Strand cDNA Synthesis Kit from Fermentas-Thermo Scientific (Burlington, ON, Canada), using random hexamers as primers, according to the manufacturer's instructions.

20 ng of cDNA were used to perform RT- or Real Time PCR amplification. For RT-PCR, template dilutions and/or differing numbers of PCR cycles were used to determine that PCR analysis was done within the linear range. PCR products were separated on a 1% agarose gel and stained with GelGreen® Nucleic Acid Gel Stain (Biotium, Fremont, CA).

Digital images were taken with the Bio-Rad ChemiDocTM Touch Imaging System and quantified using Bio-Rad Image Lab 5.2.1.

The Real Time PCR was performed using the double-stranded DNA binding dye SYBR Green PCR Master Mix (Fermentas-Thermo Scientific) on an ABI GeneAmp 7000 Sequence Detection System machine, as described by the manufacturer. The instrument, for each gene tested, obtained graphical cycle threshold values automatically.

Triplicate reactions were performed for each marker and the melting curves were constructed using Dissociation Curves Software (Applied Biosystems, CA, USA), to ensure that only a single product was amplified. Gene expression was related to 18S as housekeeping gene.

Primers sequence used for RT-PCR 13 and Real-Time PCR analyses is listed in Table 1.

### 4.7. Statistical analysis

All of data were expressed as the mean ± standard deviation (SD) of the results obtained from at least three independent experiments.

Median values, even though not shown, were calculated and analyzed to confirm significance.

Statistical evaluation was performed by one way ANOVA and Student's t-test and by Wilcoxon signed-rank test. The threshold for statistical significance was set at $p \leq 0.05$.

| RT PCR | | |
|---|---|---|
| GENE | PRIMER FORWARD | PRIMER REVERSE |
| 18S | 5'-AAA CGG CTA CCA CAT CCA AG-3' | 5'-CCT CCA ATG GAT CCT CGT TA-3' |
| KDM6B | 5'-CCTCGAAATCCCATCACAGT-3' | 5'-CCTCCAATGGATCCTCGTTA-3' |
| IL4 | 5'-TGCTGCCTCCAAGAACACAACTG- 3' | 5'-CATGATCGTCTTTAGCCTTTCCA- 3' |
| IL10 | 5' –AGATCTCCGAGATGCCTTCA- 3' | 5'-TTTCGTATCTTCATTGTCATGTA- 3' |

| REAL TIME PCR | | |
|---|---|---|
| GENE | PRIMER FORWARD | PRIMER REVERSE |
| 18S | 5'-CCC ACT CGG CAC CTT ACG-3' | 5'-TTT CAG CCT TGC GAC CAT ACT-3' |
| KDM6B | 5'-CCT CGA AAT CCC ATC ACA GT-3' | 5'-GTG CCT GTC AGA TCC CAG TT-3' |
| EZH2 | 5'-GCC AGA CTG GGA AGA AAT CTG-3' | 5'-TGT GTT GGA AAA TCC AAG TCA-3' |

**Table 1**. List and sequence of the primers used for RT-PCR and Real-Time PCR analyses.

### 4.8. Data availability

The raw data of the results presented in the paper are available upon request to the corresponding author.

### 4.9. Funding

This work was partially supported by Ethidea EXPREX project.

### 4.10. Competing interests

Dr. Massimo Balma is CEO of Ethidea Srl, the remaining authors declare that they have no competing interests.

### 4.11. Author contributions

GP carried out cellular and molecular studies and made substantial contribution to analysis and interpretation of data. GP, AF, MB and ML participated in the design and coordination of the study and drafted the manuscript. All authors have read and approved the final manuscript.

## 5. Acknowledgements

We thank Prof. Giorgio Grosa and Prof. Gianpiero Gervino for helpful discussion and comments.